# FAIR Metadata
# A Community-driven Vocabulary Application

Christopher B. Rauch[1][0000−0003−2061−3413], Mat Kelly[1][0000−0002−0236−7389],
John A. Kunze[2][0000−0001−7604−8041], and Jane Greenberg[1][0000−0001−7819−5360]

[1] Drexel University, Philadelphia PA 19104, USA
{cr625, mkelly, jg4233}@drexel.edu
https://cci.drexel.edu/mrc
[2] California Digital Library, University of California, Oakland, CA 94612, USA
jak@ucop.edu
https://cdlib.org

**Abstract.** FAIR metadata is critical to supporting FAIR data overall. Transparency, community engagement, and flexibility are key aspects of FAIR that apply to metadata. This paper presents YAMZ (Yet Another Metadata Zoo), a community-driven vocabulary application that supports FAIR. The history of YAMZ and its original features are reviewed, followed by a presentation of recent innovations and a discussion of how YAMZ supports FAIR principles. The conclusion identifies next steps and key outputs.

**Keywords:** metadata · standardization · FAIR · metadata quality · community driven

## 1 Introduction

FAIR data sharing principles have taken hold and continue to be adopted across a wide diversity of communities. While grounded in open data and data sharing, the FAIR principles further apply to sensitive data. The principles bring even more communities together to build sustainable, trusted data sharing infrastructure. Metadata is essential to FAIR, and researchers have underscored its role in key documentation [12], exemplifying what Riley calls value-added data [11]. Selected example metadata applications supporting FAIR principles include [7, 13, 5]. One area requiring further attention is the need to facilitate metadata standards development, specifically agreement on and durable reference to semantics.

In further supporting FAIR, there is a need for an open service that supports community-driven evolution, consensus-building, and permalinked references to metadata semantics currently siloed by discipline or institution type. Such an application could help communities to define, reference, and test semantics before issuing a standard based upon them. Communities could also share their semantics with other communities, learning from and borrowing prior art, avoiding duplication of effort between communities, and reducing the proliferation of metadata standards with overlapping semantics. Foundational work for this type of application was initiated with the YAMZ (Yet Another Metadata Zoo) project [4].

YAMZ was initiated as part of the NSF DataONE Datanet initiative [8] through a collaborative effort of the Metadata Preservation and Metadata Working Groups. It is important to acknowledge that YAMZ preceded the development of FAIR, although the underlying vision [4] echoes a number of foundational principles guiding today's FAIR efforts. YAMZ development is focusing on enhancing application features that integrate community member expertise to: 1) determine a canonical set of metadata terms, and 2) analyze and improve the quality of related metadata. This notion builds on the ranking/feedback loop that underlies community driven systems such as Stack Overflow [1] and Reddit [2].

This paper presents some of these YAMZ enhancements. We first review the history of YAMZ, its original features, and the initial ranking methodology. Next, we report on recent innovations and plans, followed by a discussion of how YAMZ supports FAIR principles. The paper closes with a summary of the work presented and identifies next steps.

## 2   Background: the history of YAMZ

YAMZ was developed through the efforts of the NSF DataONE Preservation, Metadata and Interoperability Working Group (PMWG). This working group was charged with developing solutions to prevalent, growing, and anticipated cyberinfrastructure preservation and metadata-related challenges. Part of DataONE's mission is to facilitate data sharing and interoperability among a diversity of earth science communities and to further bring together a growing number of community member repositories (currently 44). From the start, the DataONE PMWG took on the challenge of advancing the expensive conventional "panel of experts" process of developing shared metadata standards by turning to social computing, particularly inspired by community driven forums.

## 3   Original YAMZ

The YAMZ metadata dictionary permits any authenticated user to propose terms for inclusion with a corresponding definition and example. The user can comment on the terms submitted by other contributors, attach tags to definitions, or provide examples. During the prototype evaluation period, a sample group contributed terms to the dictionary and voted on contributed terms.

The terms are scored according to a heuristic that models consensus-based quality evaluation. Community votes influence ranking on a numerical scale intended to reflect the appropriateness or deficiency of the proposed term's definition and usage examples. Contributor reputation influences the score in that votes for terms proposed by users who have made prior contributions or comments, or that have participated in voting activity receive additional weight.

---

[1] https://stackoverflow.com/
[2] https://www.reddit.com/

The YAMZ consensus ranking system, described in detail at [9], begins with the percentage of users who cast an up-vote for a term, where $u$ is the number of up-votes and $d$ the number of down-votes. If every user casts a vote, then a term's score (S) is

$$S = \frac{u}{u+d}$$

Since not every user will cast a vote, the weight of a user's vote is adjusted by that user's reputation. Let $R_i$ be the reputation acquired by user $i$, let $R$ be the total reputation of the users who have voted on a particular term, and let $r_i = R_i/R$ for each user $i$ who voted on the term. The weight $w_i$ of a user's vote is then based on their reputation by the formula

$$w_i = 1 + r_i\,(t - v)$$

where $t$ is the total number of users in the community and $v$ the number of votes cast.

The scores are updated by a regularly scheduled process to account for changes in user reputation. Terms are assigned a stability rating depending on how long they remain unaltered by their contributor and are classified into one of three types: *vernacular, canonical,* and *deprecated.* A term is designated *vernacular* by default. After the term has exceeded a certain stability threshold (remains unaltered for an assigned duration), it becomes canonical if the consensus of the community, as indicated by the score (voting consensus), exceeds a designated threshold (75 percent). A term is deemed *deprecated* if its score (S) falls below 25 percent. These formulations are initial values and their adjustment is the subject of the ongoing enhancements as discussed in subsequent sections. Before the enhancements reported in the next section, the dictionary contained approximately 2,778 terms contributed by 158 users.

## 4 Ongoing Enhancements

This section presents YAMZ enhancements supporting metadata quality evaluation and refinement. Primarily, the proposed changes involve additional variables added to consensus ranking algorithms and the expansion of social features to facilitate community interaction. We also outline some changes in approach to the data model, such as representing subject and object terms as equivalent and relationships as predicates.

### 4.1   Equivalency of terms

The RDA Metadata Principles assert that the only difference between metadata and data is mode of use [6]. The data model of the revised metadictionary will reflect this distinction. In the proposed enhancements, a term is represented as a first-class object and corresponding examples are other terms related to that object.

Although the application of the term metadata is contextual [6], a practical distinction is possible between collections of data that exist for categorical object description and the descriptions themselves. This distinction is represented by the schematization of metadata terms by various committees or industry groups and their promulgation as standard vocabularies versus terms from those vocabularies as assigned to objects in a collection. As linked data, the resulting relationships of vocabulary terms to objects yields a set of predicates. When a set of terms is encoded according to these relationships, it can represent categorical metadata within the associated domain. By default, the categorized terms inherit the ontological status of the primary subject relative to the represented domain [10]. In the new data model, the relationships between the terms are stored separately using the RDF subject, predicate, object syntax. This supports the addition of terms organized with different schemas.

### 4.2 Scoring terms

Scoring is intended to facilitate metadata quality analysis, and the enhancements retain the ability to score terms in various contexts. The stability and appropriateness (applicability) scores encompass the metrics used to determine the category of vernacular, canonical, or deprecated assigned to a term. If a term was imported from a formal schema, the term is stable by default. When imported in this way, the source URL is recorded. When the record of a term is accessed for display, the source document is hashed and compared with a stored hash value. If the two values are the same, then stability is not affected. If they differ, the score is reduced by a set percentage, and the term is flagged. If the original document is no longer accessible, the stability score is further reduced. There are default values for the magnitude of the reduction or increase, although these may be changed in the management interface. Initial values are arbitrary but as part of a pilot project, a test group will help to adjust those weights and compare the rankings achieved by vocabularies that have been assessed for quality by a manual or semi-automated process.

When a schema is imported manually or terms are entered individually, they may languish over time as new ones become popular. Terms that are neither edited by their author nor scored by the user community lose applicability over time. Scoring can both increase or reduce the numerical quality of a term (a heuristic for appropriateness). It is increased by some amount by interaction but may decrease or increase depending on an up or down vote from a registered user and their corresponding reputation.

Retrieving lists of terms within a given stability threshold is a resource-intensive task but implementing a message queue rather than relying on database polling makes these queries feasible. Future plans include the incorporation of additional available XSEDE[3] provided resources to facilitate text analysis based

---

[3] The Extreme Science and Engineering Discovery Environment (XSEDE) is an NSF-funded organization that coordinates the sharing of digital services

on sets of terms from the dictionary and integrations with externally hosted classification tools that expose an API.

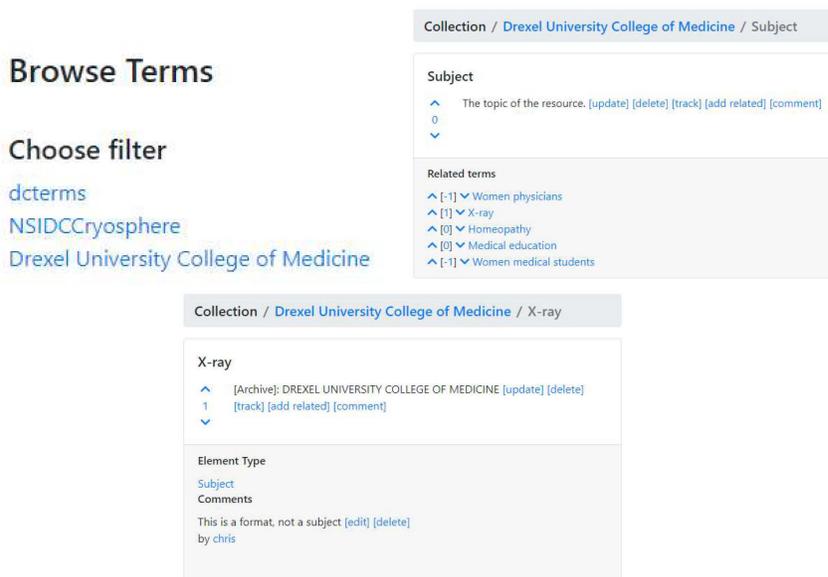

**Fig. 1.** Terms filtered by collection and refined by subject in three steps. 1. The user chooses from a list of filters. 2. The filtered vocabulary terms are presented. 3. Users vote and comment on the appropriateness of the term relative to the filters.

### 4.3   Adding and Importing terms

New subject terms can be added manually or by parsing an imported schema. Vocabulary (object terms) can be imported by parsing RDF or XML representations of records encoded according to an imported schema. Bulk import from a collection can facilitate an analysis of metadata quality if the collection is designated with a unique identifier.

The RDA Metadata Standards Catalog is an open directory of metadata standards maintained by the Research Data Alliance and hosted by the University of Bath. [4]

### 4.4   Moderation of Terms and Concepts

The originator of a vocabulary term or a set of terms imported by schema or by inference from XML or RDF records becomes the default custodian of those terms. Since anyone in the community can comment on these terms and add representative examples, initial contributions of related terms by non-custodians, opinions might differ as to the appropriateness of the related term (its fitness). While this is by design (there should be a low barrier for contributions), the custodian may assign users who are domain experts to act as moderators for a group of terms. The up and down votes of these users will be more heavily weighted for the terms they are assigned, in addition to the weight conferred by their reputations. Those users followed by the custodian will also receive additional weight when voting, but by a smaller degree than explicitly named moderators.This adjustment will allow users who have some particular domain expertise recognized by the custodian to more strongly influence the vetting of terms from a contributed vocabulary and comports with the functional principles stated above.

---

[4] https://github.com/rd-alliance/metadata-catalog-v2/blob/live/openapi.yaml

### 4.5   Sets and Surveys

Groups of terms that share some common characteristics such as schema or contributor, can be exported through the user interface as JSON or XML data and filtered or transformed for various purposes. Contributors have the ability to send surveys to their followers by notification within the YAMZ environment or by sharing a link with potential participants. These surveys can be prepared and distributed ad hoc or as part of proceeding with a specific goal, such as a metadata review session or conference.

### 4.6   User Profiles and Comments

Although users are not expected to create detailed profiles, the profile section permits linking to external services such as ORCID or LinkedIn to provide some context to the value of their contribution. Completing a profile increases the numerical value representing user reputation which will, in turn, increase the weight of a user's vote when ranking quality and, to a lesser extent, stability. The votes of more active users (those who have logged on frequently and more recently, voted or commented) will receive additional weight. Users may also follow other users, and the number of followed and followers affects reputation score.

### 4.7   Term Tracking and Notification

As in the prototype, users can elect to track terms of interest and choose to receive notifications of varying levels of activity associated with those and, optionally, related terms. An activity might include an author or administrator's update of a term (or related term), the addition of a related term, change in stability/quality within a given threshold, or a notification of a survey involving a followed term. Notifications appear in the user's profile page and can be sent by email in either a scheduled digest or as they are generated.

**Fig. 2.** User profiles display activity information, followers, contributed terms, tracked terms and users followed.

8    C. Rauch et al.### 4.8  Rights

Contributions to the YAMZ metadictionary are dedicated to the public domain under the terms of CC0 (No Rights Reserved) [1]. Imported terms are governed by the rights statements assigned by their proponent organization. Whenever possible, this is extracted from the rights statements encoding in a metadata schema, but at a minimum, a link to the source is maintained with every term entry, so that usage rights are discoverable.

### 4.9  Versioning

Updates (edits) to terms are tracked as versions, and version history can be displayed by selecting the appropriate option in the display screen of the relevant term. Currently, each discrete update is stored separately (a separate entry is created each time the term definition is edited), but versioning will eventually incorporate a more efficient approach. Terms can be submitted for review to their contributors with the posting of a comment. A list of down-voted terms and terms that have become deprecated is visible on the profile page of each user. The user is also notified by email and invited to update the entry when any change of status occurs.

### 4.10  Tagging, Searching, and External Linking

Tags are either user-created in comments, automatically assigned using a database feature that produces a list of distinct lexemes from the aggregated text of associated terms, or can be added from a by search, chosen from a list, or suggested with type-ahead [2]. Tags may also be linked to external services such as OpenRefine[5] through an API endpoint designed for that purpose.

### 4.11  Exporting Terms and Registering Vocabularies

Users can export terms individually or in sets, grouped by selectable criteria. Although not planned for this revision phase, registration and automated update of terms with various vocabulary services or semantic repositories are planned.

### 4.12  Unique Identifiers

All terms are assigned an Archival Resource Key (ARK) that uniquely identifies them and allows for their resolution (if they are resolvable) through the N2T service of the California Digital Library[6]. Imported schemas, collections, and manually added terms also receive an identifier. The ARKs carry with them persistence data that can be retrieved by adding a suffix to their URL. This data is determined from metadata relating to the terms versioning and categorization and presented as a persistence statement.

---

[5] https://openrefine.org
[6] https://n2t.net/

## 5 Discussion

Metadata scheme development and supporting metadata interoperability remain among two of the most time demanding challenges for researcher communities that seek to share, reuse, and integrate data from multiple sources. The metadata ecosystem is extensive and research points to the development of overlapping schemes that have been developed in silos. Further, there are reports of researchers spending over 80 percent of their time managing their data and less than 20 percent of their time pursuing their science. As presented above, an application like YAMZ can help address these challenges and help advance metadata activities in connection to FAIR.

### 5.1 How YAMZ Helps FAIRNness

YAMZ leverages social computing technology, the spirit of open science, and ultimately the notion of the wisdom of the crowds for metadata scheme development. The original vision that helped to shape YAMZ called for a low barrier for contributions, transparency in the review process, collective review and ownership, expert guidance, voting capacity, and overall stakeholder engagement. These aspects are reflected in FAIR principles with the support of data being findable, accessible, interoperable, and reusable. The FAIR community also recognizes the significance of vocabularies and how these principles are exhibited in metadata schemes. Cox et al. [3] identify and discuss ten rules that support FAIR vocabularies, covering governance, licensing, quality issues, persistence, and other key aspects.

The rules present a rubric for showing how YAMZ aligns with FAIR principles.

**Rules 1 and 2. (Governance)**
In YAMZ, the identification of the content custodian is multifaceted. Contributions are made under CC0 terms but rights holders of imported schemas retain ownership rights.

The unique style of governance offered by the metadictionary model that incorporates crowd-sourced ranking with the ability to import both canonical (schema elements) and vernacular (as vocabulary terms). The status of these can be adjusted by vote, comment, moderation and usage examples. YAMZ governance represents an attempt to expose as much content as possible to the public domain while recognizing that not all contributors can grant that right. Origin and commensurate publishing rights are assumed to be with a term contributor and thereby in the public domain as established by the contribution. Imported schemas will always retain a link to the original source or a stored version of the document. An effort is underway to incorporate data sharing agreements into the submissions process for proprietary vocabularies, but this is in the early planning stage.

**Rule 3 (Consistency)**
One of the goals of the project is to facilitate metadata quality analysis. Accu-racy and consistency are attributes that can be voted on and commented on. Contributors can filter a particular corpus of records representing objects imported from collections as a way to focus quality evaluation and facilitate group work. This approach works best when analyzing terms that have repositories in common, but can also be initiated as a survey with invited participants. This process facilitates consistency checking.

**Rule 4 (Maintenance Environment)**
Facilitating vocabulary maintenance is a primary goal of the project. The ranking algorithms together with the group work features and versioning capabilities ensure that changes to definitions or categorization are recorded with their associated events (either responding to a comment, author edit of recharacterization of a term by stability, category, and fitness metrics.

**Rule 5 (Unique Identifiers)**
Archival Resource Keys are appropriate identifiers for contributed terms. They are flexible enough to assign to any type of term and produce a resolvable identifier. ARKs can be minted as terms are added. When terms change, the metadata associated with object persistence that can be retrieved via an ARK inflection is updated to reflect the date and time of the change (last updated).

**Rule 6 (Machine Readable Versions)**
The dictionary will feature both REST APIs for returning data over HTTP in various formats and the ability to respond to ARK-related metadata requests. The bulk of terms will be imported and machine-readability is therefore assumed from successful import, but individual terms can also be manipulated and extracted through the API.

**Rule 7 (Vocabulary Metadata)**
The assignment of metadata is inherent in the contribution process. Provenance information includes, at the least, a record of the contributor of a term. Versioning-related metadata is recorded each time a database transaction occurs or there is user activity. Permanence statements that are retrievable by their ARK ids represent term metadata as does the association of related terms in the subject-object-predicate pattern.

**Rule 8 (Register the Vocabulary)**
The dictionary will be registered manually with various vocabulary services and semantic repositories as suggested.

**Rule 9 (Accessibility)**
The dictionary makes terms viewable by humans with an HTML representation that includes the definition with related terms. Machine access is facilitated by the API and the ability to export selected sets of terms in various formats or conventions such as JSON, XML, RDF, and SKOS.

**Rule 10 (Publishing)**
Versioning information is available at API endpoints corresponding to the following terms (at a minimum) suggested by Cox.

- **dcterms:created-date** or date-time that the vocabulary or term was initially created
- **dcterms:modified-date** or date-time that the vocabulary or term was last updated
- **dcterms:isReplacedBy** - to point to a superseding vocabulary or term
- **dcterms:replaces** - to point to a prior version of a vocabulary or term
- **owl:deprecated = 'true'** if the vocabulary or term is no longer valid
- **owl:priorVersion** - to point to a previous version of a vocabulary
- **owl:versionInfo** - general annotations relating to versioning
- **skos:changeNote** - modifications to a term relative to prior versions
- **skos:historyNote** - past state/use/meaning of a term

## 6 Conclusion

The conventional approach to developing metadata standards is slow, expensive, and compromised by egos and the limited number of people available to review, travel to, and endure endless consensus discussions. Social computing technology offers new opportunities for enabling a more inclusive, transparent approach to metadata standards development. Moreover, social computing offers community features that can be leveraged to support FAIR and reduce duplication of effort in metadata silos. The YAMZ prototype serves as a demonstration for such an approach, and increased interest in and adoption of FAIR principles present a rubric for enhancing YAMZ as described in this paper.

YAMZ was created before the articulation and adoption FAIR, but the current prototyping enhancements map well to FAIR principles and metadata rules. Next steps include demonstrating and testing these new features, following by a set of formal evaluations. The results of these activities will be folded into future releases of YAMZ.